\documentclass[%
rsi,
 amsmath, amssymb,
preprint,%
]{revtex4-1}

\usepackage{graphicx}
\usepackage{dcolumn}
\usepackage{bm}
\usepackage{caption}
\usepackage{subcaption}
\usepackage[utf8]{inputenc}
\usepackage[T1]{fontenc}
\usepackage{mathptmx}
\usepackage{etoolbox}
\usepackage{xcolor}

\makeatletter
\def\@email#1#2{%
 \endgroup
 \patchcmd{\titleblock@produce}
  {\frontmatter@RRAPformat}
  {\frontmatter@RRAPformat{\produce@RRAP{*#1\href{mailto:#2}{#2}}}\frontmatter@RRAPformat}
  {}{}
}%
\makeatother
\begin{document}

\preprint{RSI}

\title{A new compact symmetric rotational diamond anvil cell for in situ high-pressure-torsion studies}
\author{K. K. Pandey}
\homepage{kkpandey@barc.gov.in}
\affiliation{High Pressure and Synchrotron Radiation Physics Division, Bhabha Atomic Research Centre, Mumbai 400085, India}
\affiliation{Homi Bhabha National Institute, Anushaktinagar, Mumbai 400094, India }

\author{H. K. Poswal}%
\homepage{himanshu@barc.gov.in}
\affiliation{High Pressure and Synchrotron Radiation Physics Division, Bhabha Atomic Research Centre, Mumbai 400085, India}
\affiliation{Homi Bhabha National Institute, Anushaktinagar, Mumbai 400094, India }

\date{\today}

\begin{abstract}

In situ studies under severe plastic deformation at high pressures, employing rotational/shear diamond anvil cells (RDAC), have recently gained much interest in the high-pressure community owing to their potential applications in material processing methods, mechanochemistry, geophysics, etc. These studies, combined with multi-scale computational simulations, provide important insights into the transient hierarchical microstructural evolution, structural phase transitions, and orientation relationship between parent and daughter phases and help establish the kinetics of strain-induced phase transitions under severe plastic deformation. Existing RDACs are mostly used in axial x-ray diffraction geometry due to geometrical constraints providing less reliable information about stress states and texture. Their asymmetric design also poses serious limitations to high-pressure shear studies on single crystals. To overcome these limitations, a new compact symmetric rotational diamond anvil cell has been designed and developed for in situ high-pressure torsion studies on materials. The symmetric angular opening and short working distance in this new design help obtain a more reliable crystallographic orientation distribution function and lattice strain states up to a large Q range.   Here, we present the advantages of the symmetric design with a few demonstrative studies.

\end{abstract}

\maketitle

%

\section{Introduction}
Severe plastic deformation (SPD) has been in use for ages for metal forming and synthesis of high-strength materials \cite{Edalati-2022,Durand-2003,laperouse-2008,Durand-2021,Tewari-2003}. The availability of precise characterization techniques in recent decades such as scanning electron microscopy (SEM), transmission electron microscopy (TEM), electron Backscatter Diffraction (EBSD), synchrotron-based micro x-ray diffraction (XRD), etc. has made the SPD an even more effective tool for producing bulk ultrafine-grained and nanostructured materials with advanced mechanical and functional properties \cite{Valiev-2006,segal-2018}.  Depending on the methodology and complex load-shear pathways the material is subjected to, SPD can be used to introduce various kinds of lattice defects and unique microstructural features. High-pressure torsion (HPT), first introduced by Bridgmann up to a few GPa pressures \cite{Bridgman-1935,Bridgman-1952}, is one of the classic techniques of SPD used for studying grain refinement, strain hardening, structural changes, polymorphism, mechanochemistry, etc. However, most of the studies carried out using conventional tungsten carbide anvils are carried out ex-situ which is oblivious to micro-mechanical processes happening in the system while under load-shear conditions. The shear strain estimated under HPT ( $\gamma=2\pi rN/h$) \cite{Edalati-2016,Zhilyaev-2008} is based on the assumption of ideal contact friction conditions between the sample and anvils which may not be correct. The pressure estimates under HPT calculated as force/area may also be inaccurate due to the presence of heterogeneities in stress states across anvil culet. These limitations do not allow any systematic quantitative study of processes under HPT.

\begin{figure}[h!]
\includegraphics[width=0.6\linewidth]{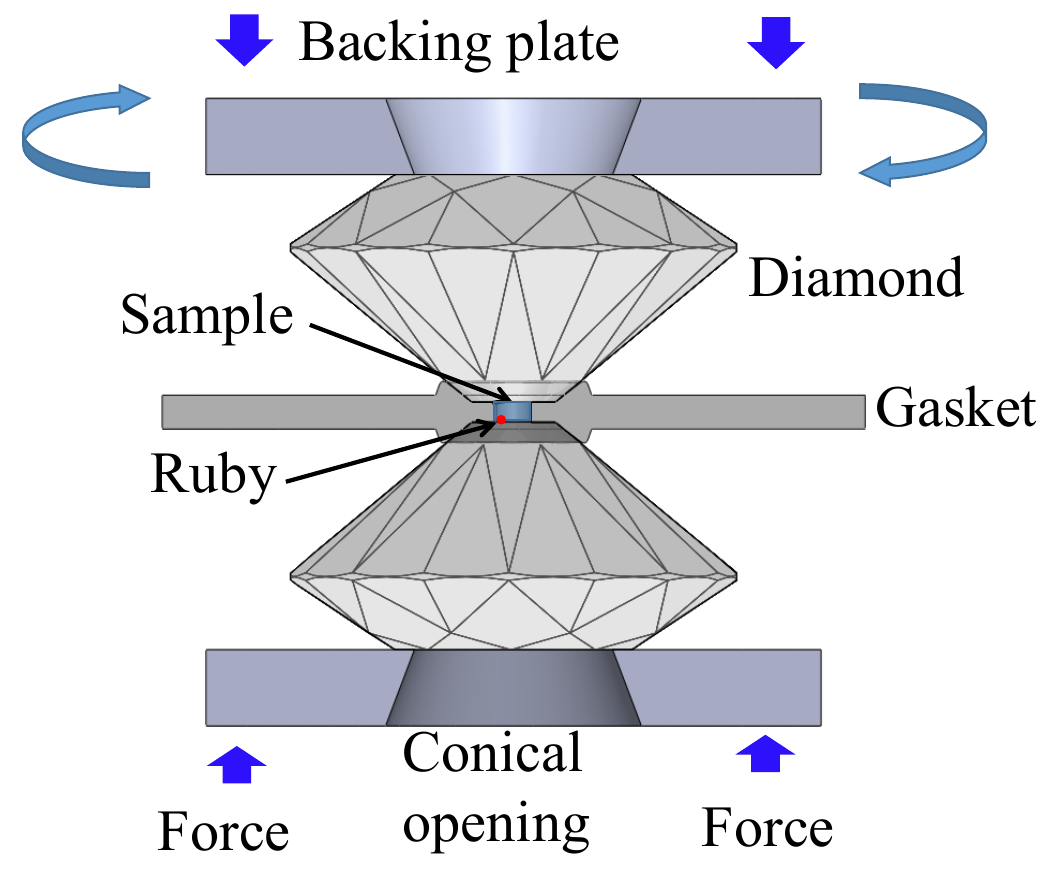}
\caption{Schematics of shear/rotational diamond anvil cell.}
\label{fig:rdacschematic}
\end{figure}

Combining the Bridgman method for high-pressure torsion and the concept of the diamond anvil cell developed in 1959 \cite{Weir-1959}, Blank et al. introduced the first shear diamond anvil cell (SDAC) in the 1980s \cite{Blank-1984, Blank-1984a} which not only provided an opportunity to increase the pressure by several tens of GPa but also facilitated in situ studies under HPT employing x-ray diffraction and Raman spectroscopy \cite{Blank-2019}. The cell was designed with an additional degree of freedom of rotation of one of the anvils with respect to another, facilitating HPT in the sample. Fig. \ref{fig:rdacschematic} shows the schematic of SDAC. Over the years, owing to the importance of in situ studies under high-pressure deformation, other research groups also developed their design of sophisticated high-pressure devices providing heterogeneous stress conditions \cite{Novikov-1999,Wang-2003, Ma-JPCS-2006,Nishihara-2008, Kawazoe-2010,Hunt-2014,Novikov-2015, Nomura-2017} and carried out numerous in situ studies on materials ranging from elemental metals, metal alloys, semiconductors to ceramics, geophysical and planetary materials bearing not only scientific but applied interest as well \cite{Edalati-2022, Mao-2016, levitas-2021,levitas-2023}. These studies help establish the classification of HP phase transitions as pressure-induced, stress-induced, or plastic strain-induced. Among these, plastic strain-induced phase transitions where plastic strain acts like a time-like parameter are especially important for in situ HPT studies \cite{levitas-2021,levitas-2023, levitas-2019}.  In most of the in situ studies under HPT in SDAC/RDAC, the primary focus has been the microstructural evolution,  polymorphism, amorphization, nature of phase transition and its mechanism, etc. with little or no emphasis on the quantitative association of plastic strains to the phase transformation kinetics.

Prof. Valery has given a multi-scale theoretical approach to explain the plastic strain-induced phase transitions and proposed a kinetic theory of these phase transitions as a function of accumulated plastic strain, initiation pressure for the phase transformation under hydrostatic conditions and complex load-shear conditions and transformed phase concentration \cite{levitas-mechchem-04,levitas-prb-2004}. The first experimental validation of this theory has been published recently only with an example study on $\alpha \rightarrow \omega$ phase transition in ultra-pure Zr \cite{pandey-acta-2020}. Such studies open a new opportunity for quantitative study of strain-induced PTs and reactions with applications to material synthesis and processing, mechanochemistry, and geophysics. 

Even though SDAC/RDAC  used for such studies \cite{Novikov-2015,Blank-2019, pandey-acta-2020} are quite sophisticated, the reliability of quantitative estimation of elastic-plastic strains, texture, etc. suffer from the technical limitations of these DACs. The XRD measurements in SDAC/RDAC are mostly carried out in axial geometry where the load axis is parallel to the incident X-ray beam. In this geometry, the diffraction condition is satisfied for the planes whose normal is mostly perpendicular to the load axis. Hence the lattice strains are under-estimated in axial geometry. To overcome this limitation, the XRD data can be recorded in radial geometry (load axis perpendicular to incident x-ray beam) \cite{Kinsland-1976,Hemley-1997,aksingh-1998} covering all stress states at different azimuthal angles of Debye rings. However, the radial geometry is out of the question here as it would provide averaged information across anvil culet diameter with no insights about the pressure profile and stress-heterogeneities in the sample across culet diameter. Recently, methodologies have been developed combining experimental observations and finite element method (FEM) to obtain more reliable stress estimates in RDAC \cite{levitas-natcomm-2023}. Another possibility is to carry out XRD measurements in oblique angle geometry (load axis at some angle between $0^\circ$ and $90^\circ$ with respect to the incident x-ray beam), however, the asymmetric design of existing SDAC/RDAC does not allow such measurements. Another disadvantage with axial geometry is the restricted coverage of pole figure \cite{Merkelwebpage} which limits the reliability of quantitative texture estimation. Besides, with asymmetric working distance at both sides, existing RDACs allow only single-sided spectroscopy measurements such as micro-Raman or Brillouin scattering measurements, that too in the backscattered geometry. Under HPT conditions, these spectroscopy measurements at both sides of RDAC become important to estimate stress states at both the sample-anvil contact surfaces. 

To overcome these limitations, a new compact symmetric RDAC (SRDAC) has been design and developed. The cell is designed with Boehler Amlax type anvils \cite{Boehler-2004} and can be used for unlimited rotations at pressures as high as 50 GPa. The pressure range may be further increased using different anvil designs and suitable backing seats. The salient features of this new design and a few example studies have been presented in the following sections to demonstrate the capabilities of the new SRDAC.

\section{Design  and salient features of SRDAC}
 
\begin{figure}[h!]
\includegraphics[width=\linewidth]{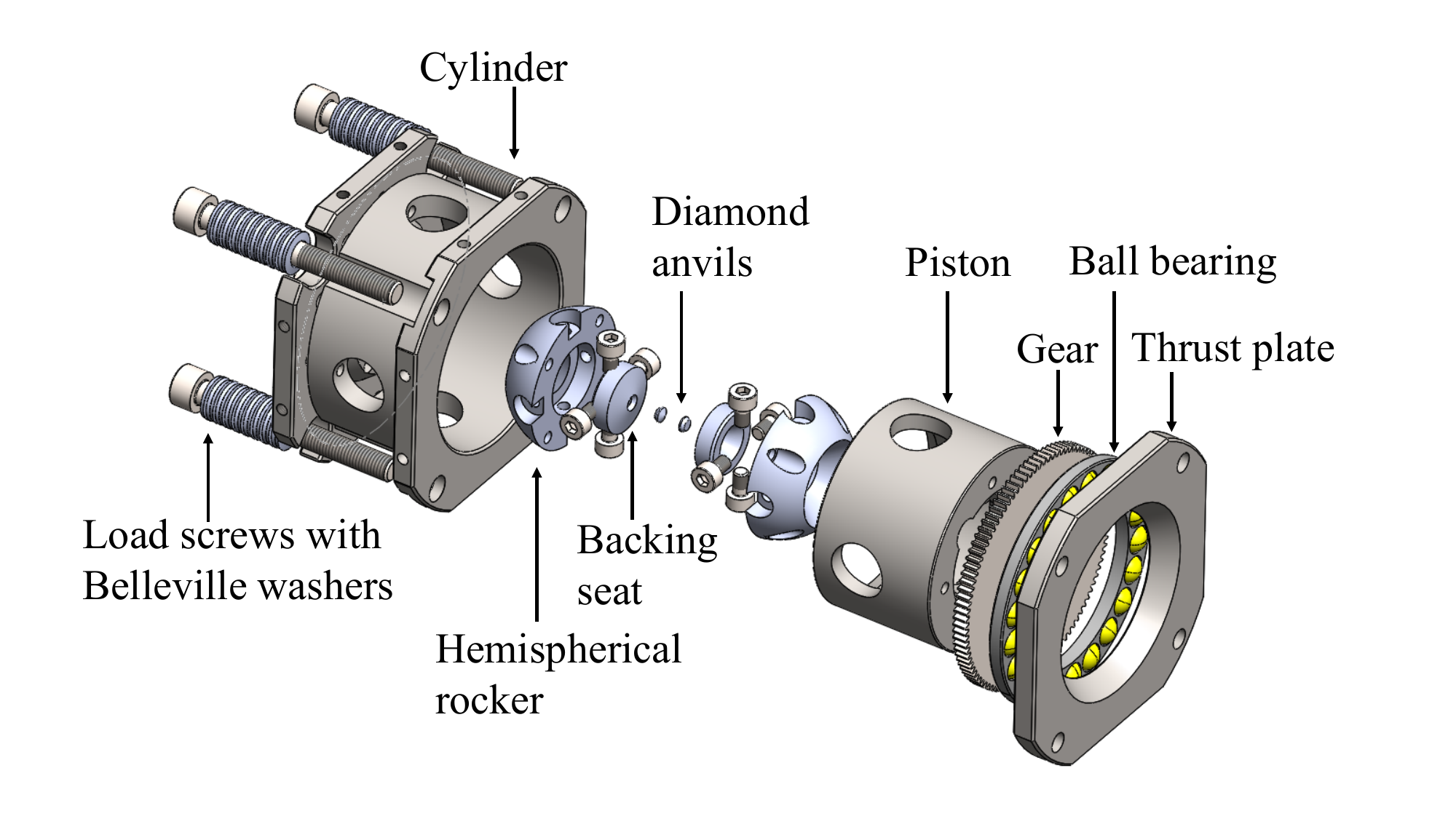}
\caption{Exploded view of the newly designed and developed compact symmetric rotational diamond anvil cell.}
\label{fig:SRDACexplodedview}
\end{figure}

The new SRDAC has been designed with symmetric angular openings on both sides with the provision of rotation of one of the diamond anvils with respect to another. As the SRDAC is to be used for in situ XRD and spectroscopic studies, special consideration is given to large symmetric angular openings, short working distance, and coaxial stability under high load/shear conditions. Simultaneous loading and rotation of the piston with respect to the cylinder requires a ball bearing between the piston and pressure plate to reduce friction. However, it poses restrictions on the working distance and angular opening of the cell. We have overcome this limitation by designing a custom ball bearing for the rotation mechanism. A piston-cylinder assembly is chosen for better translational alignment and coaxial stability of the diamond anvils during loading and rotation. For RDAC to work properly the runout of the piston should be as small as possible. We have kept it less than 5 $\mu m$. Surface roughness and waviness of the cylindrical surfaces have also been kept as small as possible. The exploded view of SRDAC design is shown in Fig. \ref{fig:SRDACexplodedview}. The SRDAC and conical diamond anvils have been manufactured by M/s Somdev Instruments Pvt. Ltd. India \cite{somdev}.
 
\begin{figure}[h!]
\includegraphics[width=\linewidth]{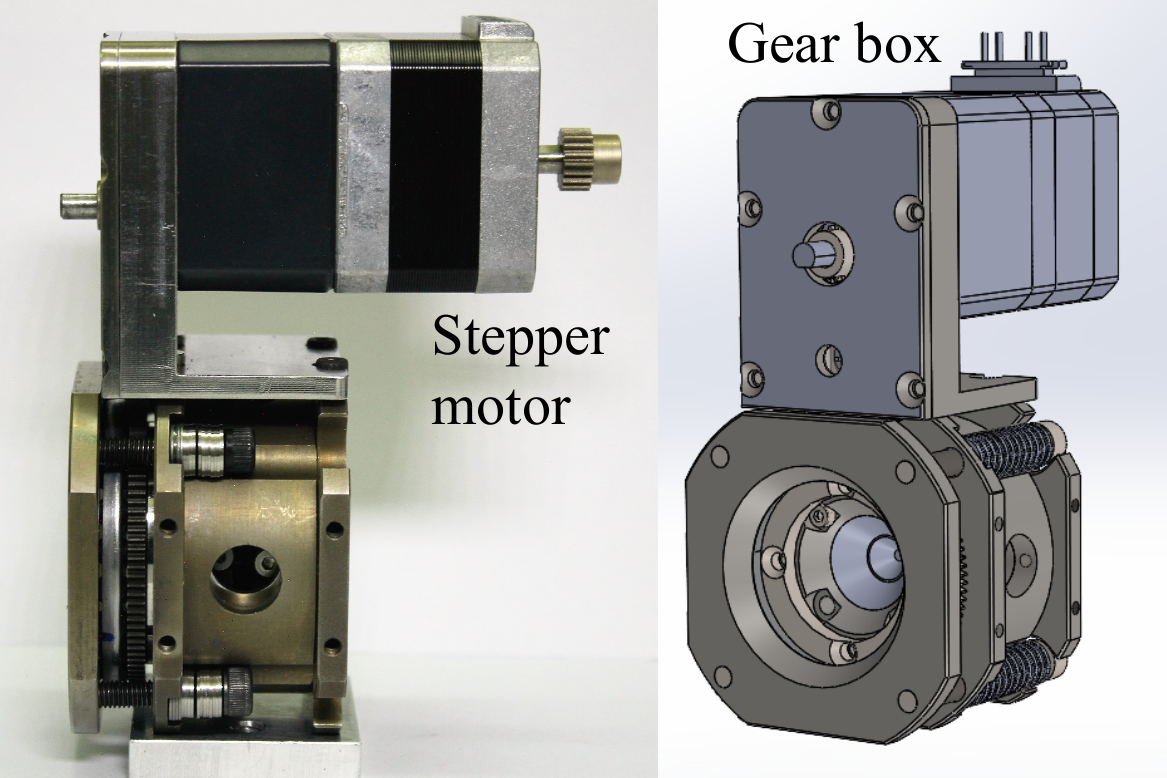}
\caption{Motorized assembly of compact symmetric rotational diamond anvil cell.}
\label{fig:SRDACfull}
\end{figure}

{\it Load and torsion mechanism:} The SRDAC is designed with an integrated gear mechanism and motorized rotation with a stepper motor. The gear mechanism rotates the anvil mounted on the piston whereas the anvil mounted on the cylinder remains fixed. There are two mechanisms for applying the load i) using four screws (two right-handed and two left-handed) which press the cylinder part against a load-applying plate with threaded holes for the screws, ii) a double diaphragm membrane for applying the load from the opposite side of the gear assembly with thrust plate. In manual loading, Belleville spring washers are used while membrane loading is done without any Belleville washers. The gear ratio between the stepper motor and anvil rotation is 1:180 which provides quasi-static shear conditions at very low-speed rotations (< 1RPM). The stepper motor with gear provides a maximum torque of $4Nm$ for the rotation of the piston which as per our experience is sufficient for the rotation of the piston anvil even at high load conditions. The design of the motorized mechanism and full assembly of SRDAC are shown in Fig. \ref{fig:SRDACfull}.

{\it Tilt and translation adjustment for both anvils:} One of the primary requirements of RDAC is to align the diamond anvils with respect to each other and with respect to the rotation axis of the RDAC. To achieve this, both the anvils are provided with tilt and translation adjustments as shown in Fig. \ref{fig:SRDACexplodedview}. The backing seat for both the anvils for the piston as well as the cylinder is designed with a hemispherical rocker for tilt adjustment and a cylindrical backing plate for Boehler-type diamond anvils, mountable within the hemispherical rockers with provision for translational alignment. Both piston and cylinder are provided with 4 fold holes for the translational alignment of anvils. 

{\it Symmetric x-ray opening:} Existing RDACs \cite{Novikov-2015,Nomura-2017,Blank-2019} are mostly piston-cylinder type where the load is applied mechanically through lever or screw mechanism. Due to the technical requirements, these RDACs have asymmetric X-ray openings at the piston and cylinder sides. In order to carry out oblique angle XRD measurements, the RDAC needs to be tilted with respect to the incident x-ray beam, however, very limited angular opening at the piston side ($\sim 2^\circ$) does not allow these measurements. Though, with a large lateral opening, high-pressure rotational deformation apparatus developed by Nomura et al. \cite{Nomura-2017} can be used for oblique angle XRD and x-ray laminography, the spatial resolution of stress heterogeneities within the sample may be compromised. We have overcome these limitations by designing an RDAC with a symmetric opening on both sides. This new design provides a compact symmetric opening of $60^\circ (4 \theta)$ at both sides, facilitating oblique angle XRD measurements from $-30^\circ$ to $+30^\circ$. With the symmetric X-ray opening, the SRDAC can also be used for single-crystal XRD measurements under high-pressure torsion. 
  
%

{\it Compact design:} 
The SRDAC has been designed to be as compact as possible without compromising its capabilities. To have a compact size of SRDAC and good stability, the piston cylinder engagement is maximized by keeping the anvil mounting assembly inside the piston. The overall diameter of the piston-cylinder is $\sim 53 mm$.  The working distance at both sides of the piston-cylinder assembly is $<20 mm$ on both sides using objectives of outer diameter $\sim 26 mm$ such as Olympus SLMPlan 20x.  This facilitates spatially resolved micro-Raman studies across samples from both sides. This is also useful for recording high spatially-resolved ruby fluorescence images at both sides of the sample implementing recently developed displacement field methodology \cite{pandey-jap-2021}. With these measurements, the plastic flow of the sample at the sample-anvil contact surface and actual torsion in the sample can be obtained quantitatively. The overall weight of the entire assembly including the stepper motor for rotation and the gear mechanism is $\sim 1kg$. The footprint of DAC is $\sim 60mm x 50mm$. With the compact design, the SRDAC can be easily carried and mounted at different laboratory-based experimental facilities as well as synchrotron facilities.

 \begin{figure}[h!]
\includegraphics[width=\linewidth]{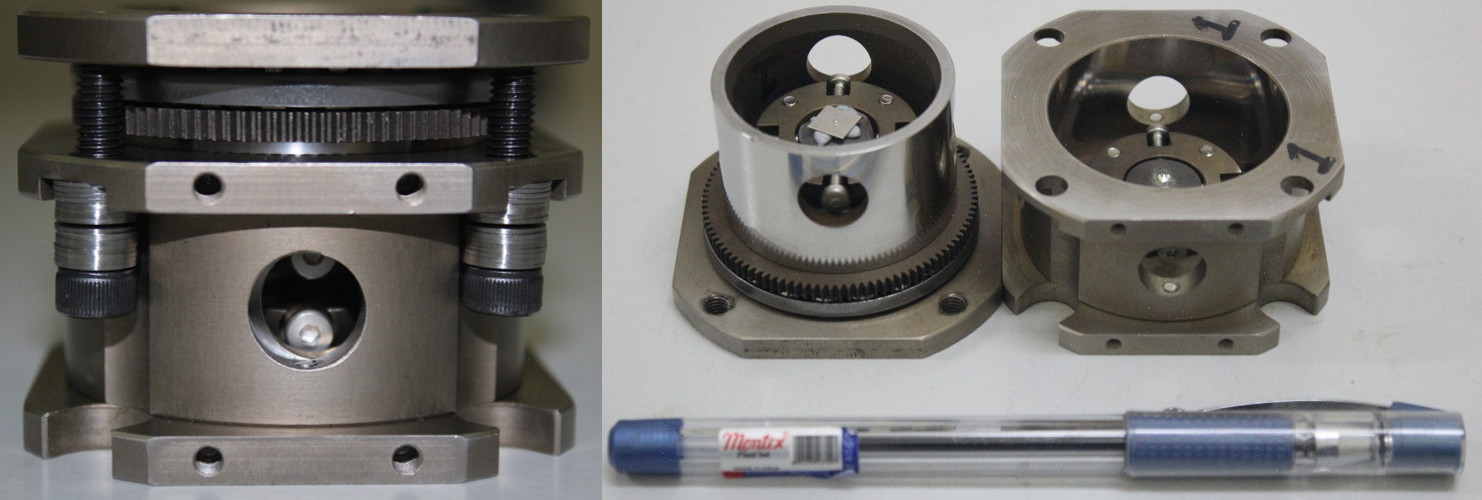}
\caption{Piston and cylinder of SRDAC.}
\label{fig:SRDACpistoncylinder}
\end{figure}

\section{Alignment and testing}
\begin{figure}[h!]
\includegraphics[width=0.6\linewidth]{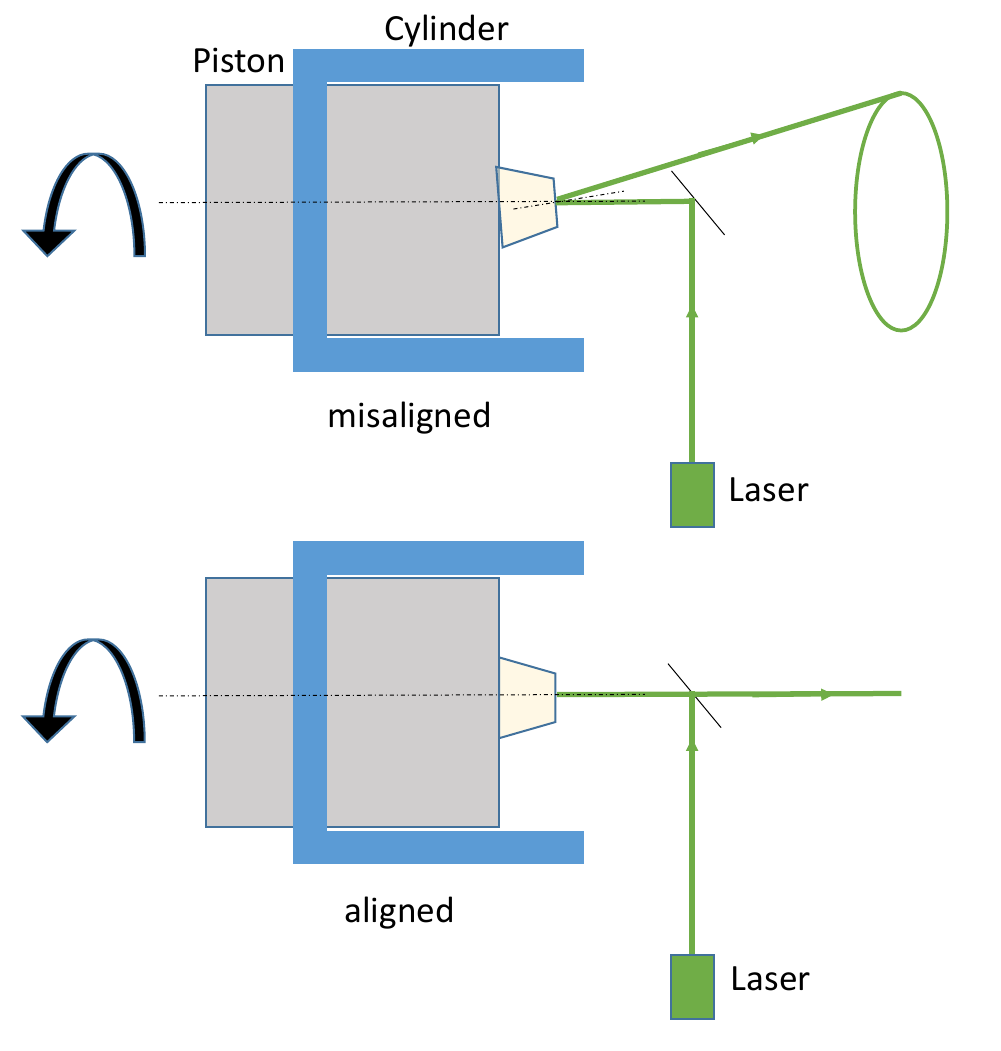}
\caption{Schematic of arrangement for anvil alignment.}
\label{fig:anvilalignment}
\end{figure}

The anvils of SRDAC need to be aligned with respect to each other and the rotation axis of the piston-cylinder assembly of SRDAC. For this, first, the anvil was mounted on the piston side only and it was aligned with respect to the rotation axis using a monocular zoom microscope and laser reflection method as shown in Fig. \ref{fig:anvilalignment} Once this anvil is aligned, the other anvil is mounted on the cylinder side and aligned with respect to the already mounted anvil under a microscope using interference fringe method.

\begin{figure}[h!]
\includegraphics[width=0.9\linewidth]{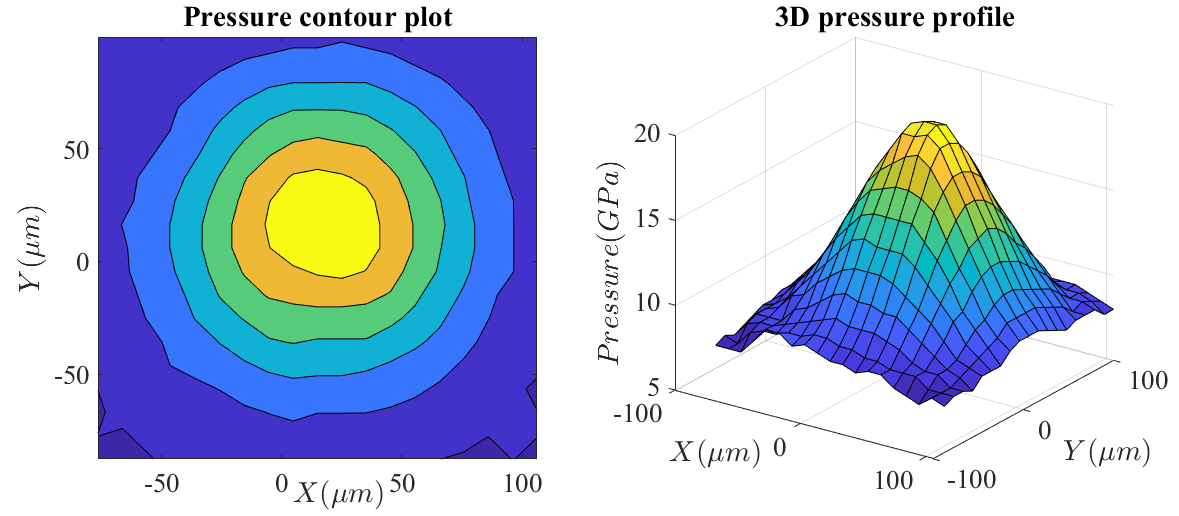}
\caption{Pressure distribution in Ni gasket at a given load condition in SRDAC.}
\label{fig:pressuredistribution}
\end{figure}

To confirm the stability of alignment under load conditions, a Ni gasket was loaded in SRDAC along with several ruby particles of size $\sim 2 - 5 \mu m$ at the sample-anvil contact surface and was subjected to uniaxial compression. At the given load condition, the pressure distribution in SRDAC was estimated using the ruby fluorescence method as well as the shift in the diamond Raman peak. As shown in Fig. \ref{fig:pressuredistribution} the pressure distribution shows a symmetric conical shape which confirms the alignment of anvils under high load conditions. 
  
\section{Demonstrative Studies}

\subsection{Oblique angle XRD measurements}
\begin{figure}[h!]
\includegraphics[width=0.8\linewidth]{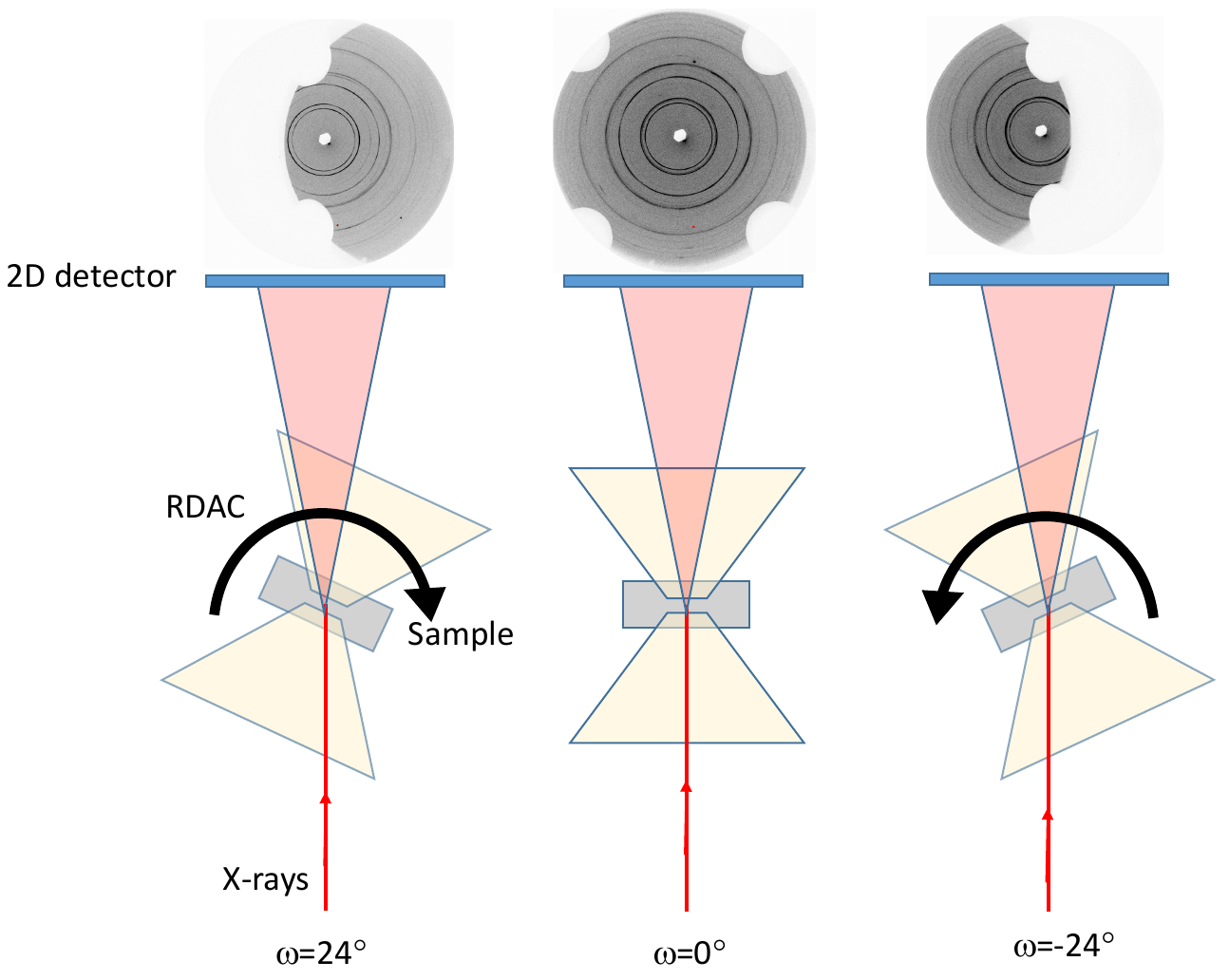}
\caption{Schematic of oblique angle XRD measurements in SRDAC.}
\label{fig:obliquegeometry}
\end{figure}

\begin{figure}[h!]
\includegraphics[width=0.8\linewidth]{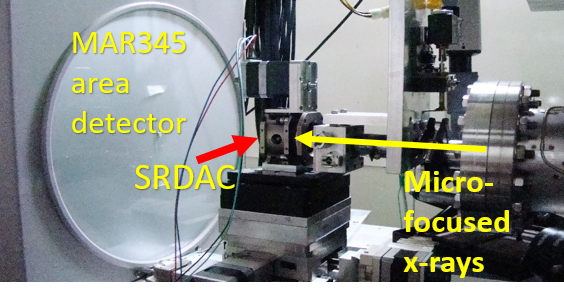}
\caption{Experimental station at the Extreme conditions XRD (ECXRD) beamline, BL-11 at Indus-2 synchrotron source, India \cite{BL-11} with SRDAC mounted at the sample stage.}
\label{fig:BL11withSRDAC}
\end{figure}

With symmetric X-ray openings at both sides, the newly developed SRDAC can be used for oblique angle XRD measurements, providing more reliable stress estimates besides capturing stress heterogeneities and pressure profiles across anvil culet. The geometry of oblique angle XRD is shown in Fig. \ref{fig:obliquegeometry}. The symmetric XRD opening of $\sim \pm 30^\circ$ in the new design of SRDAC allows oblique angle XRD measurements up to $\sim 28^\circ$. To compare the lattice strain estimation in axial and oblique angle measurements, a Zr sheet of thickness $\sim 100\mu m$ was loaded in SRDAC and subjected to some uni-axial load. At the same load condition, XRD measurements were carried out in axial geometry and at oblique angles of $24^\circ$. The measurements were carried out using micro-focused monochromatic X-rays ($\lambda$ =0.6649 \AA ;FWHM = 10 $\mu m$) at Extreme conditions XRD (ECXRD) beamline, BL-11 at Indus-2 synchrotron source, India \cite{BL-11} (Fig.\ref{fig:BL11withSRDAC}). 

\begin{figure}[h!]
\includegraphics[width=\linewidth]{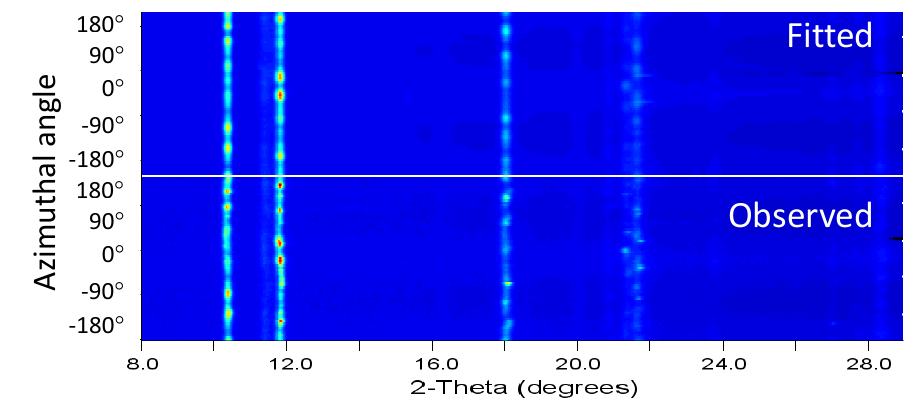}
\caption{Observed and fitted cake integration of XRD image of $\alpha$-Zr recorded in axial geometry. }
\label{fig:axialcakeintegeration}
\end{figure}

\begin{figure}[h!]
\includegraphics[width=\linewidth]{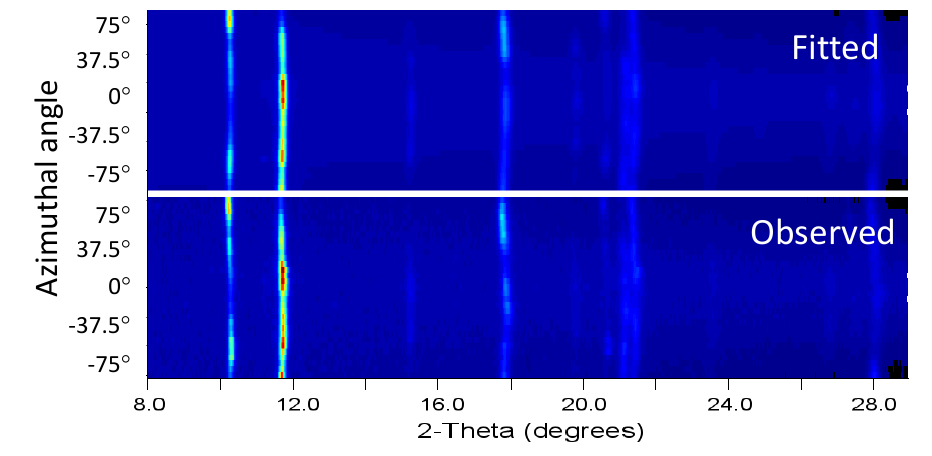}
\caption{Observed and fitted cake integration of XRD image of $\alpha$-Zr recorded in oblique geometry. }
\label{fig:obliquecakeintegeration}
\end{figure}

To maintain the same sample-to-detector distance, the sample was aligned at the rotation axis of the sample stages at the beamline which is pre-aligned with respect to the incident X-rays.  The 2D  diffraction images in these geometries were integrated into several azimuthal angle sectors of $5^\circ$ and analyzed using MAUD software \cite{MAUD}. As can be seen in the fig. \ref{fig:axialcakeintegeration} and fig. \ref{fig:obliquecakeintegeration}, the cake integration of diffraction peaks in axial geometry is almost straight implying the same lattice strain in all azimuthal directions, whereas, for oblique angle XRD, the diffraction peaks exhibit curvature which is a signature of different lattice strains at different azimuthal directions. 

To estimate deviatoric stresses at the center, the macro stress components ($\sigma_{11},\sigma_{22},\sigma_{33}$) were refined using the tri-axial model as implemented in MAUD software. The $\sigma_{33}$ is along the load axis and $\sigma_{11}$ and $\sigma_{22}$ are perpendicular to the load axis, in the $0^\circ$ azimuthal plane and perpendicular to this plane, respectively. For axial geometry thus obtained macro-stress components are found to be very close to zero ($\sigma_{33}=-0.003 \pm 0.001$ GPa, $\sigma_{11}=\sigma_{22}=0.0015 \pm 0.001$ GPa) which is not consistent with the expected macro-stresses during plastic deformation in SRDAC. In the plastic deformation regime in SRDAC, the deviatoric stress should correspond to the yield strength of the material as $\sigma_{y}=| \sigma_{33}-\sigma_{11}|$. For oblique geometry the macro-stress components are found to be $\sigma_{33}=-0.8 \pm 0.008$ GPa and $\sigma_{11}=\sigma_{22}=0.4 \pm 0.008$ GPa. Here we have used the constraint $\sigma_{11}=\sigma_{22}=-0.5\sigma_{33}$ which is a valid assumption at the symmetry axis of SRDAC.  Thus obtained yield strength of $\alpha-$ Zr is 1.2 GPa at $\sim 2$ GPa, which is consistent with the earlier reported values \cite{levitas-natcomm-2023}.  Hence, oblique geometry XRD measurements, possible with our newly designed SRDAC, provide more reliable quantitative estimates of macro stresses as compared to the axial geometry. These macro-stress components can be used to estimate the pressure-dependent yield strength of materials. It is worth mentioning here that even oblique angle XRD measurements would give information averaged over the oblique thickness of the sample parallel to the x-ray beam. However as sample thickness ($h$) in RDAC is usually small ($ < 50 \mu m$) and oblique angle ($\omega$) is $24^\circ$, the averaging is over a length scale of $htan\omega$ which turns out to be $< 25 \mu m$.  In the future, new methodologies combining experiments and macro-FEM simulations will be developed similar to Ref. \cite{levitas-natcomm-2023} for a more precise quantitative estimate of stress states and their heterogeneities with oblique angle XRD measurements.
 
\subsection{Texture studies with higher pole figure convergence}
\begin{figure}
     \centering
     \begin{subfigure}[b]{0.45\linewidth}
         \centering
         \includegraphics[width=\textwidth]{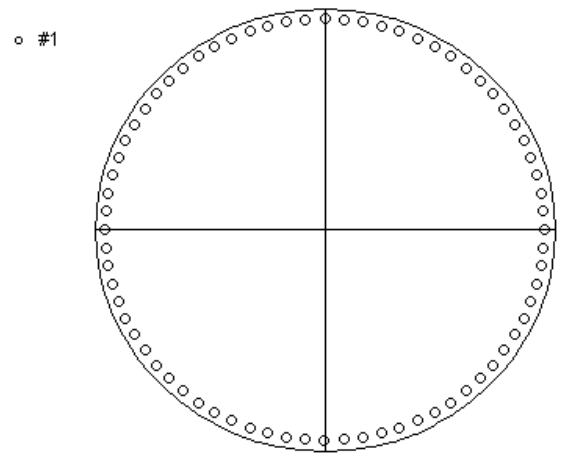}
         \caption{}
         \label{fig:polefigconvaxial}
     \end{subfigure}
     \hfill
     \begin{subfigure}[b]{0.45\linewidth}
         \centering
         \includegraphics[width=\textwidth]{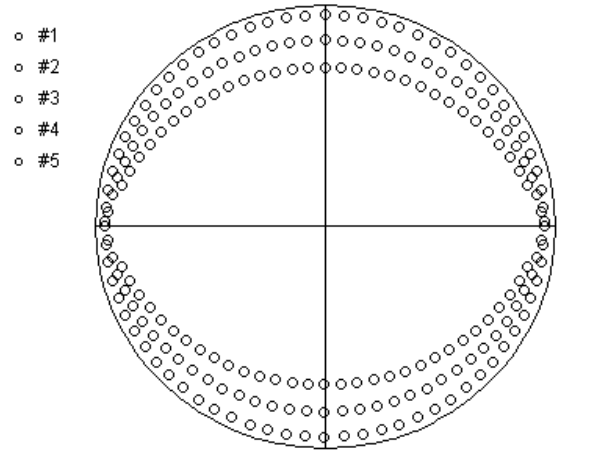}
         \caption{}
         \label{fig:polefigconwithallomega}
     \end{subfigure}
     
        \caption{(a) Pole figure convergence for XRD data recorded in axial geometry; (b) Pole figure convergence for XRD data recorded at different $\omega$ angles ranging from $-24^\circ$ to $24^\circ$ about the axial geometry..}
        \label{fig:polefig}
\end{figure}

The study of texture evolution under plastic straining provides vital insights into the deformation mechanism and in the case of phase transitions, it helps establish the orientation relationship between parent and daughter phases. For quantitative studies, the accuracy of texture parameters is important which depends on the pole figure convergence in the XRD measurements. In axial geometry, the pole figure convergence is limited as shown in Fig. \ref{fig:polefigconvaxial}. 

\begin{figure}[h!]
\includegraphics[width=\linewidth]{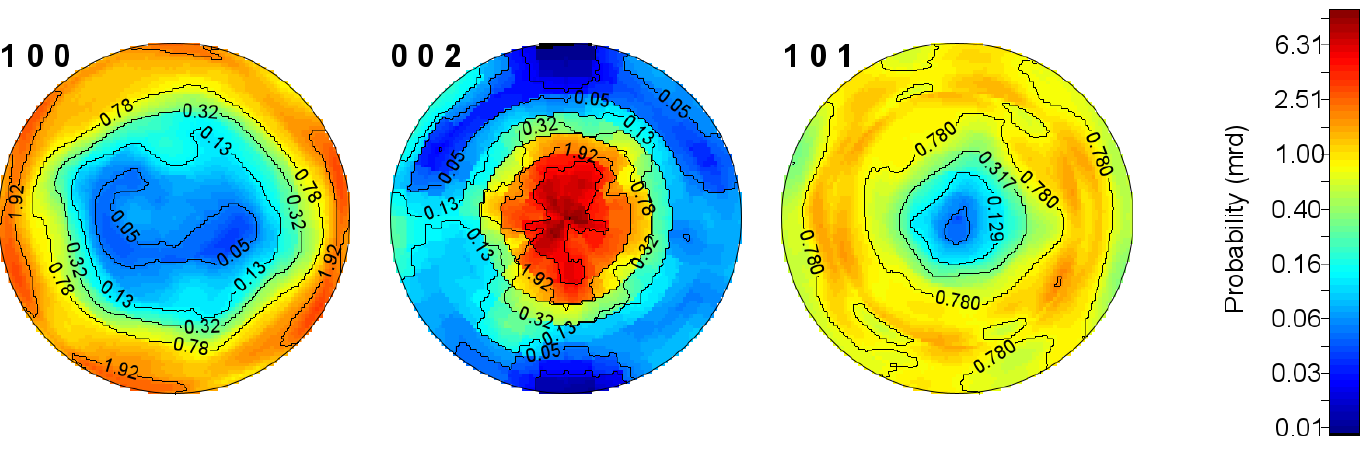}
\caption{Reconstructed pole figure of (100), (002), and (101) diffraction planes using XRD data recorded at different $\omega$ angles about the axial geometry.}
\label{fig:polefigure}
\end{figure}

\begin{figure}[h!]
\includegraphics[width=0.6\linewidth]{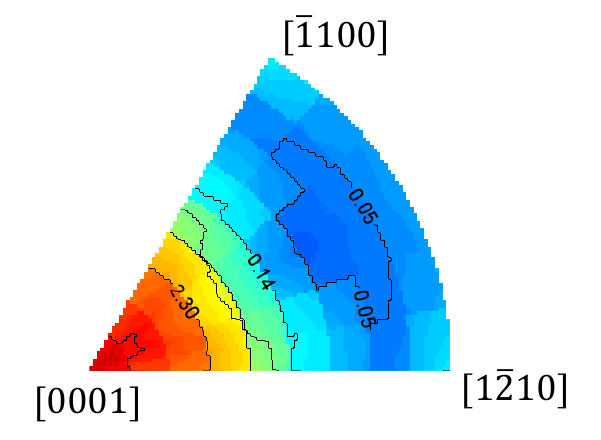}
\caption{Inverse pole figure obtained from texture analysis exhibiting strong preferred orientation along [0001] direction.}
\label{fig:inversepolefigure}
\end{figure}

The pole figure convergence can be improved by recording XRD images at different $\omega$ angles. With the new design of SRDAC, XRD images can be recorded at different $\omega$ angles as far as geometrical constraints allow. As a demonstrative study, we recorded XRD images of the Zr sample loaded in SRDAC with $\omega$ angles ranging from $-24^\circ$ to $+24^\circ$ in steps of $4^\circ$. Fig. \ref{fig:polefigconwithallomega} shows the pole figure convergence using all the diffraction data recorded at different $\omega$ angles. The quantitative texture analysis has been carried out using these diffraction data employing the E-WIMV texture model as implemented in MAUD software. Fig. \ref{fig:polefigure} and \ref{fig:inversepolefigure} show the reconstructed pole figures of a few diffraction planes and inverse pole figures respectively. The analysis shows the strong preferred orientation of (002) planes along the load axis. The texture has nearly fiber symmetry which is expected, even though the analysis has been done without imposing any symmetry on the texture model.

\subsection{Displacement field measurements}
For displacement field measurements in DAC/RDAC, a digital image correlation-based methodology has been recently developed \cite{pandey-jap-2021}. These measurements are very important to obtain adhesion zones and relative sliding between sample and anvil which can be used as the boundary conditions for finite element method simulations of processes in DAC/RDAC instead of hypothetical friction conditions. In this method, several ruby particles of size $\sim$ 2 to 5 $\mu m$ are placed at the sample anvil contact surface at both sides of DAC, and magnified ruby fluorescence images are recorded from these ruby particles before and after load/shear conditions. These images are analyzed using the digital image correlation method to find out the displacement field in DAC/RDAC. For better accuracy of measurements, the resolution of fluorescence images should be as good as possible. 

\begin{figure}
     \centering
     \begin{subfigure}[b]{0.45\linewidth}
         \centering
         \includegraphics[width=\textwidth]{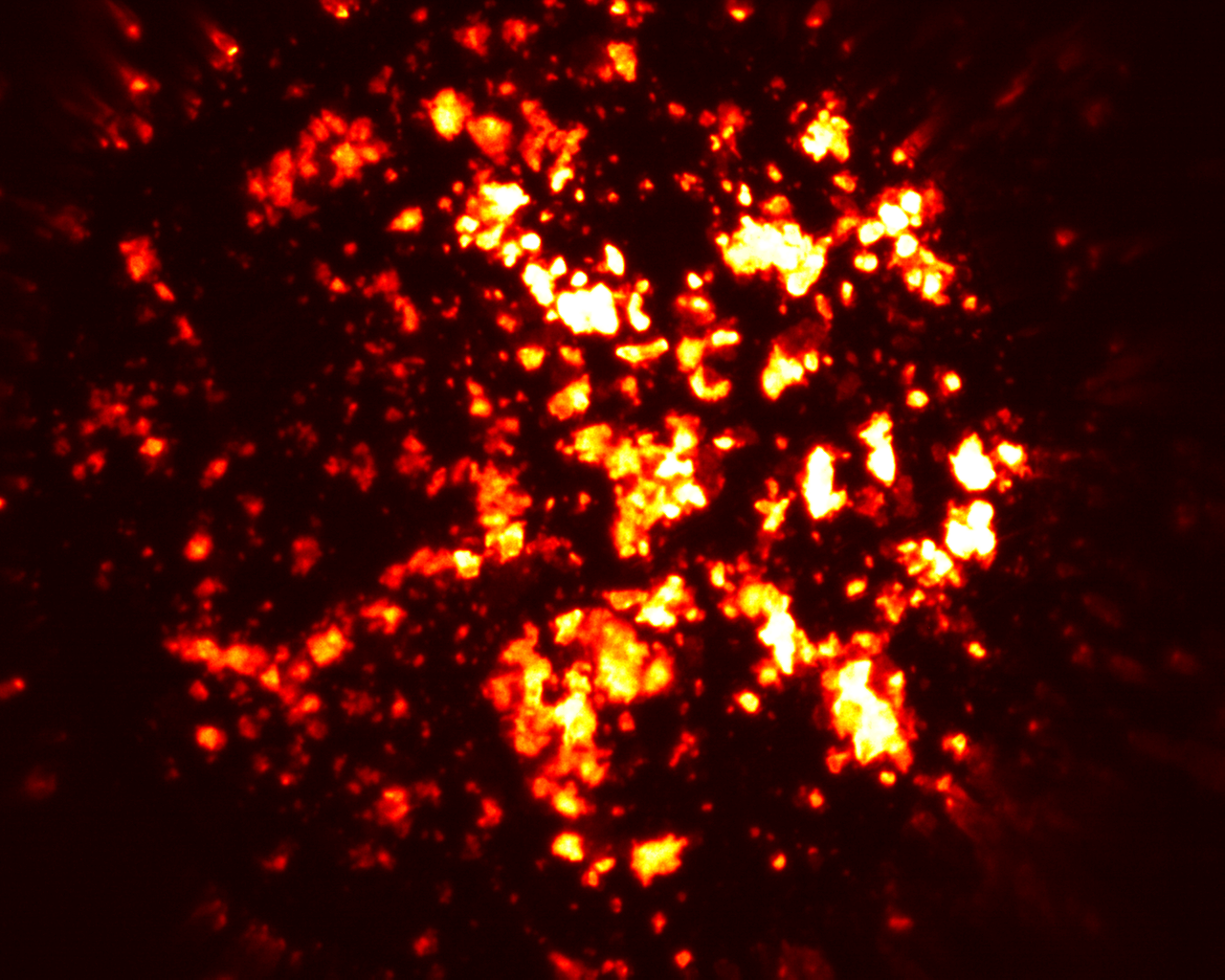}
         \caption{}
         \label{fig:shortfocal}
     \end{subfigure}
     \hfill
     \begin{subfigure}[b]{0.45\linewidth}
         \centering
         \includegraphics[width=\textwidth]{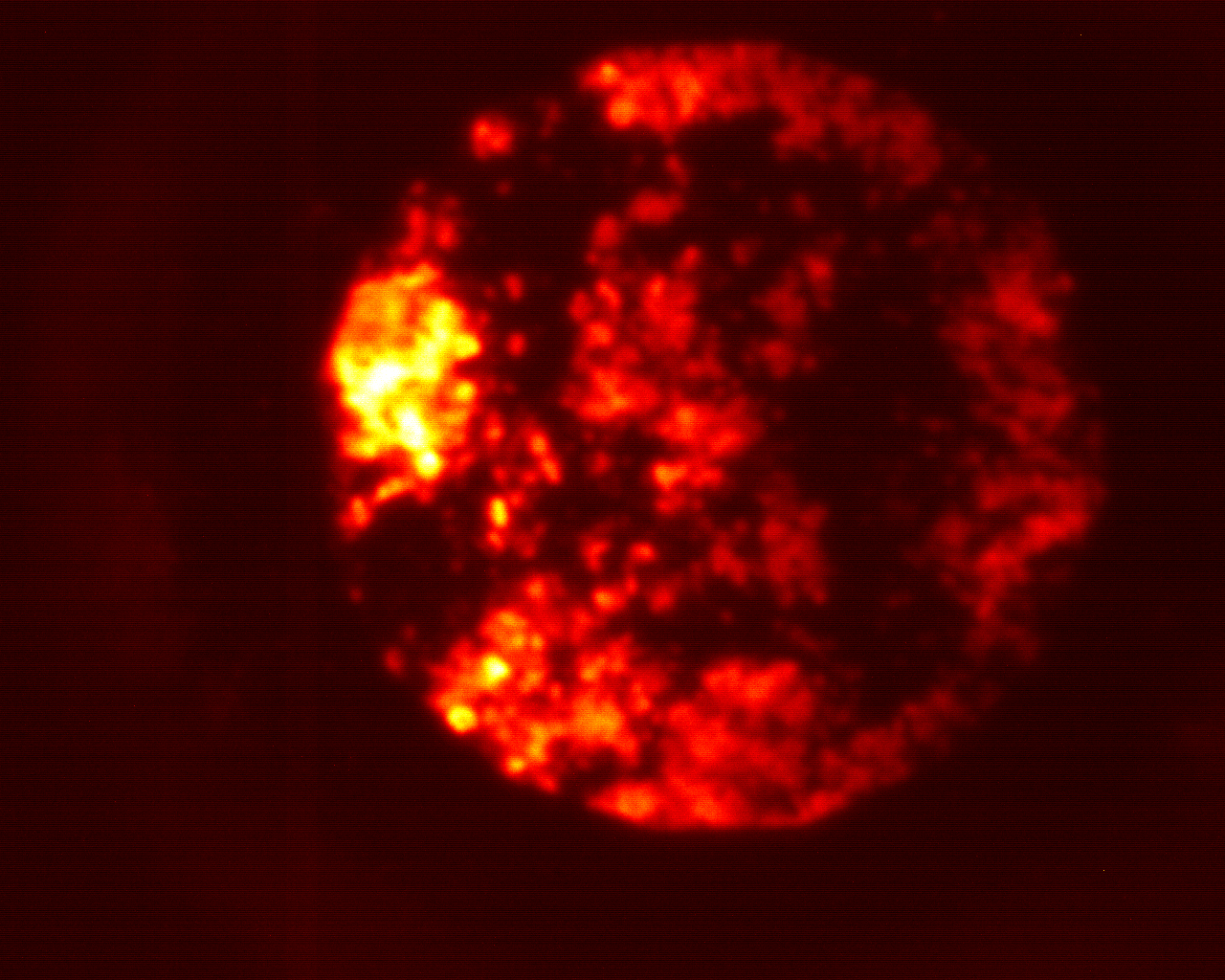}
         \caption{}
         \label{fig:longfocal}
     \end{subfigure}
     
        \caption{Ruby fluorescence image recorded from SRDAC with an objective focal length of  (a) 35 mm and (b) 120 mm.}
        \label{fig:rubyimage}
\end{figure}

The optical setup for getting these images has to be designed for both sides as per the available working distance. At best the resolution of the image is diffraction-limited and depends on the numerical aperture (NA) of the optics. For better resolution, the optical aperture should be as large as possible, and the focal length of the objective lens should be as small as possible. Existing designs of SDAC/RDAC are asymmetric with longer working distance (typically > 100 mm) and smaller aperture (typically <10 mm) at one side (piston side) which limits the resolution of the image to $\sim 7\mu m$ for the spectral range of ruby fluorescence. Whereas at the cylinder side, short working distance allows standard objectives with NA $\sim 0.28$ (focal length:35 mm) providing lateral resolution of $\sim 1.2 \mu m$.  Fig. \ref{fig:shortfocal} and  \ref{fig:longfocal} show ruby fluorescence images recorded with optical setup of focal length 35 mm and 120mm respectively which demonstrates the difference in resolution.    
The new compact symmetric design of SRDAC has a working distance of $\sim$ 20 mm at both sides and $30^\circ$ aperture. Hence, with optimal optical design for ruby fluorescence image collection setup, one may get high-resolution fluorescence images at both sides of SRDAC for more accurate estimates of displacement fields. 

\section{Conclusions}
With symmetric angular opening and short working distance at both sides, the newly designed compact SRDAC  facilities implement recently developed methodologies for more accurate quantitative studies of plastic strain-induced phase transitions. The oblique angle XRD, possible with the new design, provides more accurate estimates of stress states under HPT. Besides, quantitative texture estimates are also more accurate with larger convergence of pole figures by collecting XRD data at different $\omega$ angles. With shorter and symmetric working distance, the SRDAC  can be used for spatially resolved Raman mapping and displacement field measurements at both sides, providing more insight into the processes happening under high-pressure torsion.  The compact size of SRDAC is easy to carry and use at laboratory-based as well as synchrotron experimental facilities. 

\begin{acknowledgments}
We acknowledge the institutional support for this development work. We also acknowledge Mr. M. M. Tandel, Dr. Velaga Srihari, Dr. K. Phaneendra, Dr. M. Modak and Dr. Abhilash Dwivedi for user support at the ECXRD beamline (BL-11), Indus-2. We also acknowledge Mr. A. K. Poswal for his valuable suggestions and for critically reviewing the design of the cell. Author, K. K. Pandey would like to acknowledge Prof. Valery I. Levitas, ISU, USA for introducing him to the research field of plastic strain-induced phase transitions during his post-doctoral tenure at ISU, USA.
\end{acknowledgments}

\section*{AUTHOR DECLARATIONS}
\subsection*{Conflict of Interest}
The authors have no conflicts to disclose.
\subsection*{Author Contributions}
{\bf Dr. K. K. Pandey:} Conceptual design (equal); detail mechanical design and development (supporting); methodology (lead); experimental investigations and analysis (lead); writing-reviewing \& editing (lead). {\bf Dr. H. K. Poswal} Conceptual design (equal); detail mechanical design and development (lead); investigations (supporting), writing-reviewing \& editing (supporting).

\section*{Data Availability Statement}
The data presented in this manuscript is available from the corresponding author upon reasonable request.

\end{document}